# A Cryogenic Dielectric Antenna for Wireless Sensing and Interfacing Outside the 10 K Environment

Ingrid Torres[1], and Alex Krasnok[1,2*]

[1]Department of Electrical and Computer Engineering, Florida International University, Miami, Florida 33174, USA

[2]Knight Foundation School of Computing and Information Sciences, Florida International University, Miami, FL 33199, USA

*To whom correspondence should be addressed: akrasnok@fiu.edu

**ABSTRACT**

The performance and scalability of cryogenic microwave systems, particularly for quantum processors, are fundamentally limited by the thermal stability and loss of their constituent dielectric materials. While mixed titanate ceramics like $MgTiO_3$–$CaTiO_3$ (MCT) and $(Zr,Sn)TiO_4$ (ZST) are primary candidates, their comparative performance as radiative antennas in the deep-cryogenic regime has remained uncharacterized. Here we present a side-by-side comparison of MCT and ZST operated as dielectric resonator antennas from 296 K down to 7–10 K under identical fixtures and protocols. While the MCT resonator exhibits large, nonlinear frequency drift (~230 MHz by 10 K), pronounced thermal hysteresis, and a collapse of the loaded quality factor at low temperature—behavior consistent with incipient/relaxor-like losses, the ZST resonator demonstrates exceptional stability. Its resonant frequency shifts by only ~30 MHz, its loaded Q-factor is enhanced by ≈20–25%, and it shows negligible thermal hysteresis. Leveraging these properties, we operate the ZST disk as a radiative antenna at 10 K with only 1 mW input, establishing a through-window wireless link that detects room-temperature dielectric targets over multiple wavelengths via near-field frequency shifts and far-field magnitude modulations. This presents a viable path toward non-invasive cryogenic diagnostics and wireless interconnects that circumvent the thermal load of physical cabling. Our findings establish ZST as a foundational material for high-coherence quantum interfaces and provide a practical template for designing wireless cryogenic systems.

## 1. INTRODUCTION

Cryogenic microwave systems are an enabling technology for a host of frontier scientific and technological domains, from radio astronomy to deep-space communications (*1–5*). Operation at temperatures approaching 4 K provides fundamental performance advantages by profoundly suppressing Johnson-Nyquist noise and minimizing intrinsic material dissipation. This yields a dramatic enhancement in a component's quality (Q) factor, defined as the ratio of stored energy $U$ to dissipated power $P_{\text{loss}}$ per cycle: $Q = \omega_0 \frac{U}{P_{\text{loss}}}$, where $\omega_0$ is the angular resonant frequency. The resulting high-Q components are essential for detecting faint signals from the cosmic microwave background (*6–8*), maintaining high-fidelity data links over astronomical distances (*9*), and enabling precision measurements in quantum sensing (*5, 10*).

Perhaps the most demanding application is in scalable quantum computing. Protecting the fragile coherence of superconducting qubits for high-fidelity gate operations and quantum non-demolition (QND) readout requires their operation in an ultra-low-noise environment below 100 mK (*11*, *12*). As these quantum processors grow, they necessitate an increasing number of microwave I/O channels, leading to a critical engineering challenge known as the "interconnect bottleneck" (*13*, *14*). Conventional coaxial cables introduce a parasitic heat load that consumes the limited thermal budget of the cryostat's mixing chamber, constraining system scale. Consequently, developing wireless links to transmit signals to and from the cryogenic stage without physical tethers is a critical research thrust, offering a transformative path toward scalable and thermally efficient quantum architectures (*15*).

Dielectric resonators (DRs), which confine electromagnetic energy via the excitation of specific Mie modes (e.g., $TE_{01}\delta$), are ideal candidates for realizing such wireless cryogenic interfaces (*16–18*). Their high permittivity allows for compact device footprints, while their low intrinsic loss enables material Q-factors that can exceed $10^6$. When designed for efficient coupling to free space, they function as high-performance dielectric resonator antennas (DRAs), forming the basis of a potential wireless link (*19*, *20*). The thermal stability of a DR is governed by its temperature coefficient of resonant frequency ($\tau_f$), which is intrinsically linked to material properties through the relation:

$$\tau_f = \frac{1}{f_0}\frac{\partial f_0}{\partial T} \approx -\left(\frac{1}{2}\tau_{\epsilon_r} + \alpha_L\right) \quad (1)$$

where $\tau_{\epsilon_r}$ is the temperature coefficient of permittivity and $\alpha_L$ is the linear thermal expansion coefficient. The choice of dielectric material that dictates these parameters is therefore paramount.

Two mixed-titanate systems are of particular interest. The $MgTiO_3$–$CaTiO_3$ (MCT) composite is engineered to achieve $\tau_f \approx 0$ at room temperature by composing materials with opposing signs of $\tau_{\epsilon_r}$ (*21*). However, the incipient ferroelectricity of the $CaTiO_3$ component (*22*) suggests this delicate balance may degrade at low temperature via nonlinear permittivity and relaxor-like domain dynamics. In contrast, the $(Zr,Sn)TiO_4$ (ZST) solid solution is designed for intrinsic thermal stability and low loss (*23–25*). A critical open question is how the material physics underpinning these room-temperature designs behaves in the deep-cryogenic regime.

Despite their importance, a direct, side-by-side comparison of these materials *as radiative antennas* under identical deep-cryogenic conditions has been absent from the literature. In this work, we address this gap through a comprehensive analysis of MCT and ZST disk resonator antennas from 296 K down to 7 K. We quantify their resonant performance and investigate the low-temperature breakdown of the material engineering principles described by (1). Critically, we demonstrate that the high-stability ZST resonator can be successfully operated as a radiative antenna, establishing a functional wireless link from the 10 K environment through a cryostat window. These results provide a quantitative benchmark for material selection and, more importantly, offer a proof-of-concept for the wireless architectures needed to overcome the interconnect bottleneck in next-generation cryogenic systems.

## 2. RESULTS AND DISCUSSION

## A. Electromagnetic Design and Simulated Performance

Full-wave electromagnetic simulations were performed using CST Microwave Studio to establish a theoretical performance baseline and validate the experimental design, which is conceptually illustrated in Fig. 1. For both resonators, the fundamental transverse electric mode, $TE_{01\delta}$, was targeted. This mode is highly desirable for cryogenic applications as its field configuration—characterized by a purely azimuthal electric field and axial/radial magnetic field components—maximizes energy storage within the high-permittivity dielectric. This confinement minimizes radiation leakage, a primary loss channel, thereby enabling a very high intrinsic or unloaded quality factor ($Q_u$) (26). Simulations used open (add space/PML) boundaries, a hexahedral adaptive mesh, and a single-ended wave port referenced at the fixture feed plane; material models were set to frequency-independent $\epsilon_r$ and $\tan\delta$ consistent with room-temperature literature values, i.e., no temperature dependence was included in EM runs.

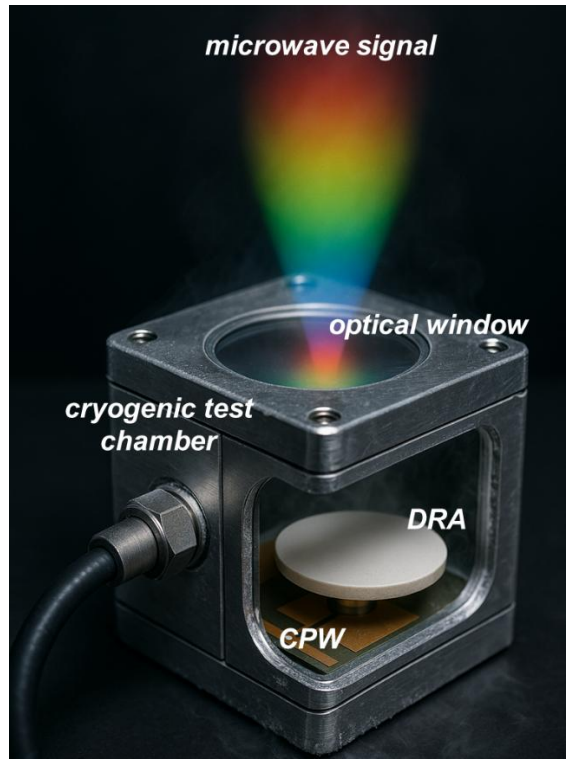

**Figure 1**. *Conceptual illustration of the cryogenic dielectric resonator antenna (DRA). The ZST ceramic disk, mounted on a coplanar waveguide (CPW) feedline within a cryogenic test chamber, functions as an efficient antenna. It radiates a microwave signal (artistically represented by the visible spectrum) through an optical window (not included in the EM radiation pattern simulations), establishing a wireless link from the cold interior to the external, room-temperature environment.*

For the $MgTiO_3$–$CaTiO_3$ (MCT) disk (nominal $\epsilon_r \approx 20$), free-space simulations confirmed the fundamental $TE_{01\delta}$ mode at 3.81 GHz (Fig. 2a). When placed in the measurement configuration atop a CPW and ground plane, the resonant frequency shifts to 3.41 GHz due to loading effects from the substrate and ground plane. The simulated reflection coefficient ($S_{11}$) exhibits a minimum of –5.92 dB at this frequency (Fig. 2b). The simulation revealed strong magnetic field confinement within the disk (Fig. 2c), quantified by an H-field enhancement factor of approximately 33 when the frequency is at 3.38 GHz (defined here as $\max|H|/\max|H|_{\text{feed}}$). This parameter is critically important for quantum circuit applications, as the coherent coupling strength $g$ between a resonator and a superconducting qubit (e.g., a transmon) scales with the resonator's zero-point magnetic field amplitude $B_{\text{rms}}$ ($g \propto B_{\text{rms}}$) (27, 28). A strong field enhancement is therefore a prerequisite for enabling fast, high-fidelity quantum gate operations and is a key metric for resonator design in circuit quantum electrodynamics (cQED). Using the one-port relation $|S_{11}(f_0)| = |(1-\beta)/(1+\beta)|$, the simulated on-resonance $|S_{11}| = 10^{-5.92/20} \approx 0.506$ implies $\beta \approx 0.33$ (undercoupled), consistent with the shallow dip.

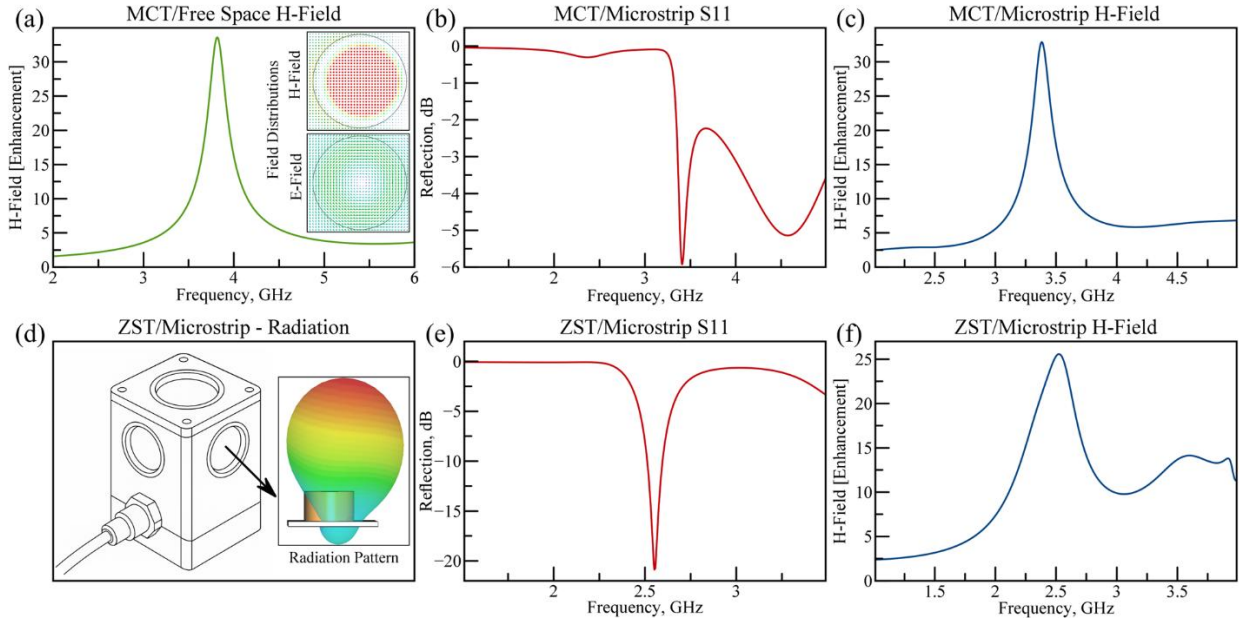

*Figure 2. Simulated electromagnetic performance and design of the MCT and ZST dielectric resonators. (a) Simulated free-space magnetic-field (H-field) spectrum for the MCT disk, showing the fundamental $TE_{01}\delta$ mode at 3.81 GHz. Insets show the corresponding H-field and E-field distributions. (b) Simulated reflection coefficient ($S_{11}$) for the MCT disk coupled to a coplanar waveguide (CPW), exhibiting a resonance at 3.41 GHz with a minimum −5.92 dB, implying an under-coupled condition $\beta \approx 0.33$. (c) H-field enhancement for the coupled MCT resonator (defined as $max|H|/max|H|_{feed}$), peaking at ~33 for 3.38 GHz, indicating strong magnetic-energy confinement. (d) Schematic of the ZST resonator in the test fixture and its simulated far-field radiation pattern, revealing a forward-biased main lobe aligned with the CPW/window normal; patterns were computed with the fixture ground plane present and the cryostat window excluded. (e) Simulated $S_{11}$ for the coupled ZST configuration, showing a deep resonance at 2.55 GHz with $|S_{11}|$=−20.88 dB, corresponding to near-critical coupling $\beta \approx 0.835$. (f) H-field enhancement for the coupled ZST resonator (same definition as in (c)), peaking at ~26 for 2.52 GHz, confirming strong field–matter interaction potential.*

A parallel analysis for the $(Zr,Sn)TiO_4$ (ZST) disk (nominal $\epsilon_r \approx 38$) predicted its fundamental $TE_{01}\delta$ mode at 2.80 GHz in free space. In the coupled configuration, the resonance occurred at 2.55 GHz with a significantly deeper $S_{11}$ minimum of –20.88 dB (Fig. 2e). The depth of this reflection null provides insight into the coupling condition of the resonator. For a one-port resonator, the minimum $S_{11}$ value (in dB) at resonance is related to the internal loss rate ($\kappa_{int}$) and the external coupling rate ($\kappa_{ext}$) by (29):

$$S_{11,\text{min}}[\text{dB}] = 20\log_{10}\left|\frac{\kappa_{int} - \kappa_{ext}}{\kappa_{int} + \kappa_{ext}}\right|. \quad (2)$$

The deep null for ZST indicates that the system is operating near the critical coupling condition ($\kappa_{int} \approx \kappa_{ext}$), where power transfer to the resonator is maximized. This suggests that the ZST material model has intrinsically lower losses (a smaller $\kappa_{int}$) and is better impedance-matched than MCT, a prediction to be tested by cryogenic measurement. With $|S_{11}| = 10^{-20.88/20} \approx 0.090$, we obtain $\beta \approx 0.835$, i.e., near-critical coupling with efficient power transfer, consistent with the deep simulated dip. The ZST resonator exhibited a similarly strong H-field enhancement factor of approximately 26 when the frequency is at 2.52 GHz (Fig. 2f), confirming its suitability for applications requiring strong field-matter interaction.

The far-field radiation patterns were simulated to evaluate the resonators' performance as antennas and to inform the design of the wireless cryogenic link. The key performance metric for this application is directivity (D), which quantifies the antenna's ability to concentrate radiated power in a specific direction. It is formally defined as: $D = \frac{4\pi U_{max}}{P_{rad}}$, where $U_{max}$ is the maximum radiation intensity and $P_{rad}$ is the total radiated power. The MCT resonator, with its lower permittivity, exhibited a quasi-omnidirectional pattern with a peak directivity of 5.12 dBi. While consistent with a symmetric $TE_{01}\delta$ mode, this broad radiation is suboptimal for a point-to-point wireless link, as significant power is radiated in untargeted directions.

In contrast, although the ZST resonator's simulated peak directivity is 4.27 dBi (numerically lower than MCT), its higher permittivity and interaction with the ground plane produce a more forward-biased beam and a larger front-to-back ratio (5.54), which better aligns the main lobe with the window normal and reduces spillover into the cryostat (Fig. 2d). This directive profile is therefore advantageous for a through-window wireless link, improving coupling to the external receiver/target and mitigating multipath and unintended radiative heating of cryogenic components.

**B. Cryogenic Performance of the MgTiO$_3$–CaTiO$_3$ (MCT) Resonator**

The temperature-dependent performance of the MCT resonator was characterized from 296 K down to 7 K, with the reflection spectra ($S_{11}$) shown in Fig. 3a. The data reveal a material fundamentally unsuited for stable, high-coherence cryogenic applications due to three critical factors: a large and non-linear frequency drift, a collapse in the quality factor at low temperatures, and pronounced thermal hysteresis.

At 296 K, the fundamental mode was measured at 3.38 GHz. Upon cooling, the resonant frequency ($f_0$) shifted significantly upwards. Between 296 K and 150 K, a range where the resonance remains well-defined, the frequency increased by 80 MHz to 3.46 GHz. The temperature coefficient of resonant frequency ($\tau_f$) over this range is calculated to be approximately -160 ppm/K, following:

$$\tau_f = \frac{(f_f - f_i)}{f_{\text{avg}}(T_f - T_i)}$$

This substantial drift accelerates at lower temperatures, with the total frequency shift reaching approximately 230 MHz by 10 K ($f_0 \approx 3.61$ GHz), corresponding to an average $\tau_f$ of -230 ppm/K over the entire range. This behavior is a direct consequence of the material's composite nature, as described by (1). The positive and highly non-linear temperature coefficient of permittivity ($\tau_{\epsilon_r}$) of the $CaTiO_3$ component dominates at low temperatures. As an incipient ferroelectric, the permittivity of $CaTiO_3$ is described by the Barrett formula (*30*), which predicts a steep increase as temperature approaches 0 K, overwhelming the balancing effect of $MgTiO_3$ and leading to the large, unstable frequency shift observed. We note that (1) is a linearized relation near room temperature; the observed cryogenic behavior reflects departure from linearity.

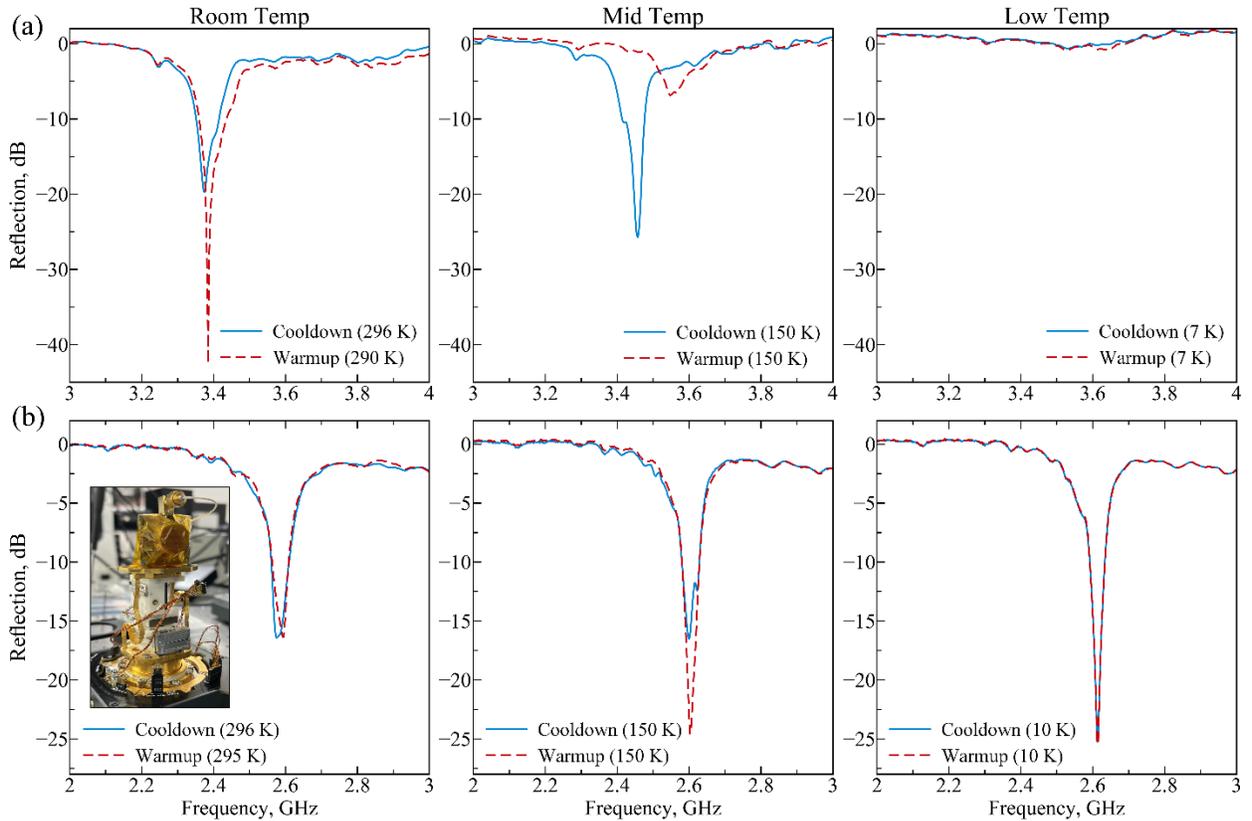

*Figure 3. Experimental cryogenic performance comparison of MCT and ZST resonators. (a) Measured temperature-dependent reflection ($S_{11}$) spectra for the MCT resonator, revealing significant resonant frequency drift, a collapse of the quality factor at 7 K, and pronounced thermal hysteresis between the cooldown (solid blue lines) and warmup (dashed red lines) cycles. At the matched 150 K setpoint, the hysteresis in frequency is $\Delta f_0^{hyst} \approx +150$ MHz (warmup–cooldown). (b) Measured $S_{11}$ spectra for the ZST resonator over the same temperature range, demonstrating excellent thermal stability, a modest Q-factor enhancement at 10 K, and negligible hysteresis. Inset: Photograph of the cryogenic measurement setup showing*

*the resonator mounted on the cryostation's sample stage. Identical VNA settings and reference plane were used across all temperatures.*

The loaded quality factor ($Q_L$), a measure of the resonator's total losses, follows a non-monotonic and ultimately detrimental trend. The total dielectric loss can be modeled as the sum of intrinsic and extrinsic contributions, $tan\delta = tan\delta_{int}(T) + tan\delta_{ext}$. From 296 K to 150 K, $Q_L$ shows a modest increase from 34 to 37. This is consistent with the expected suppression of intrinsic, phonon-scattering-related losses ($tan\delta_{int}$) upon cooling. However, below 150 K, the performance degrades sharply. The resonance broadens and shallows until it is nearly indistinguishable from the noise floor at 7 K, where $Q_L$ collapses to just 7, a degradation of 79.1%. This indicates the activation of a new, dominant extrinsic loss channel at low temperatures. A plausible mechanism is relaxor-like domain-wall motion in the $CaTiO_3$ fraction, a well-known and highly dissipative process that can dominate microwave losses at cryogenic temperatures (*31*). While our data are consistent with this picture, we refrain from a definitive assignment without direct domain-dynamics measurements.

Finally, the MCT resonator exhibits severe thermal hysteresis, as is evident in the 150 K data in Fig. 3a. The resonant frequency during the warmup cycle is approximately 150 MHz higher than during the cooldown cycle at the same temperature setpoint. Defining $\Delta f_0^{hyst} = f_0^\uparrow - f_0^\downarrow$, the nonzero value indicates path dependence consistent with slow relaxation of polar nanodomains. For any precision application, this is a critical issue: the operating frequency would depend on thermal history, complicating calibration and multiplexing. In summary, despite the strong simulated magnetic-field enhancement, the combination of large $|\tau_f|$, $Q_L$ collapse at low $T$, and thermal hysteresis makes the present MCT composition/fixture unsuitable for frequency-critical cryogenic resonators; materials/process optimization aimed at suppressing relaxor behavior would be required before reconsideration for such use.

## C. Cryogenic Performance of the (Zr,Sn)TiO₄ (ZST) Resonator

In stark contrast, the ZST resonator displayed the exemplary stability and performance enhancement required for high-coherence systems, as shown in Fig. 3b. Over the entire 296 K to 10 K temperature range, the resonant frequency shifted by only 30 MHz (from 2.58 GHz to 2.61 GHz). This corresponds to a small and stable average $\tau_f$ of -40.4 ppm/K, confirming that the material's engineered thermal compensation, described by (1), remains robust down to deep-cryogenic temperatures. We use $\tau_f$ as an average over the measured range.

Crucially, the resonator's quality factor improved significantly upon cooling. The loaded Q-factor ($Q_L$) is determined by the combination of the resonator's internal losses ($Q_u$, the unloaded Q-factor) and its external coupling ($Q_c$) via $1/Q_L = 1/Q_u + 1/Q_c$. Experimentally, $Q_L$ increased from 27.6 at 296 K to 33.9 at 10 K, a 22.6% enhancement. Using the on-resonance relation for a one-port resonator $|S_{11}(f_0)| = |(1-\beta)/(1+\beta)|$ with $\beta \equiv Q_u/Q_c$, the measured dips $-16.5$ dB (296 K) and $-25.3$ dB (10 K) yield $|S_{11}| = 10^{-16.5/20} \approx 0.1496$ and $10^{-25.3/20} \approx 0.0543$, hence $\beta_{296} \approx 0.740$ and $\beta_{10} \approx 0.897$. From $Q_u = (1+\beta)Q_L$, we obtain $Q_u^{296} \approx 48.0$ and $Q_u^{10} \approx 64.3$, confirming that the intrinsic (unloaded) Q improves upon cooling. Equivalently, the decay-rate components (reported as $\kappa/2\pi = f_0/Q$) change from $\kappa_{int}/2\pi \approx 53.7$ MHz (296 K) to 40.6 MHz (10 K), while the external rate evolves from 39.8 MHz to 36.4 MHz, indicating movement toward near-critical coupling consistent with the deeper dip. The total linewidth narrows from $\kappa_L/2\pi = f_0/Q_L \approx 93.5$ MHz to 77.0 MHz (≈ −18%), matching the observed increase in $Q_L$. At the lowest temperatures, the Q-factor is expected to plateau, limited by temperature-independent extrinsic factors such as oxygen vacancies, impurities, and grain boundaries rather than phonon losses; thus, the cryogenic $Q$ serves as a sensitive probe of microstructural and chemical purity.

The ZST resonator exhibited no discernible thermal hysteresis; cooldown and warmup traces at matched setpoints overlapped within our measurement repeatability (Fig. 3b), demonstrating that the material state is thermodynamically stable and fully reversible, a hallmark of a reliable component for precision instrumentation. For fixed external coupling, the observed $Q$ increase implies a proportional reduction of the resonator decay rate $\kappa$, which is advantageous for narrow-band, multiplexed readout and minimizes frequency collisions/crosstalk as systems scale. Together—low $|\tau_f|$, improved $Q_u$, and negligible hysteresis—these results identify ZST as a robust baseline dielectric for deep-cryogenic resonators and dielectric resonator antennas.

**D. Demonstration of a Wireless Cryogenic-to-Room-Temperature Link**

Leveraging its exceptional cryogenic stability and directive radiation pattern, the ZST resonator was successfully operated as a radiative antenna to establish a functional wireless link. This experiment was designed to demonstrate a proof-of-concept for non-invasive sensing, transmitting a signal from the 10 K stage through the cryostat and detecting a target in the external, 296 K environment. Such a capability is a critical step toward developing wireless interconnects that can circumvent the thermal and mechanical challenges of physical cabling in complex cryogenic systems. Hereafter, we denote the baseline with the Antenna Window Open as AWO; the "window closed" trace quantifies the additional standing-wave error introduced by the sealed aperture.

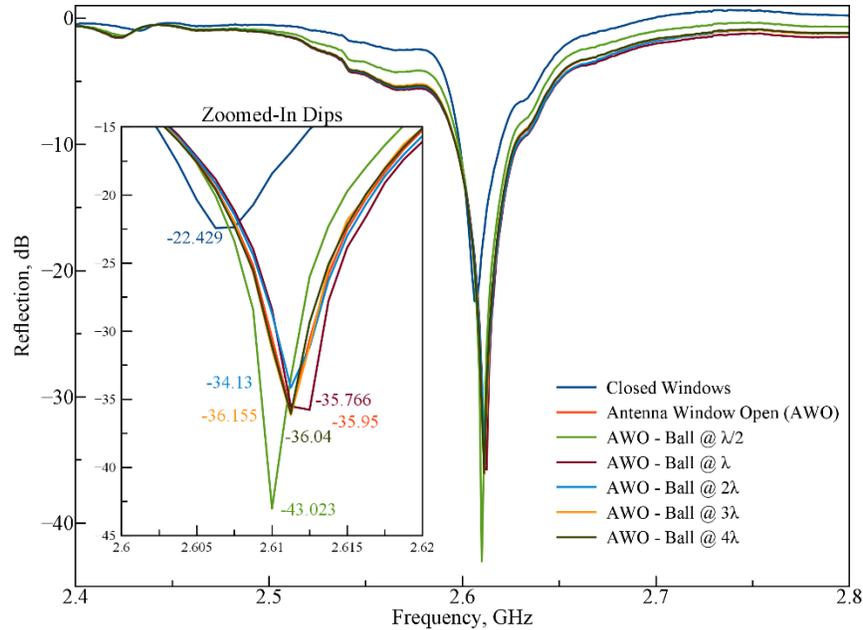

*Figure 4. Remote sensing performance of the ZST resonator antenna at 10 K. Measured reflection ($S_{11}$) spectra showing the antenna's response to a dielectric sphere placed at various distances $d$ from the cryostat window. The baseline measurements with the window closed and open (AWO) are compared to measurements with the target placed at multiples of the operating wavelength $\lambda$. The inset quantifies the dip depths: the "window closed" baseline is shallower ($-22.43$ dB), while the AWO baseline is $\approx -35.95$ dB; introducing the target at $\lambda/2$ deepens the dip to $\approx -43.02$ dB, with other distances producing dips in the $-34$ to $-36$ dB range. These controls show that the window itself introduces a standing-wave structure, and that our analysis references change relative to the AWO baseline. A clear response is observed even at $4\lambda$, consistent with monostatic $1/d^4$ attenuation yet detectable owing to the narrow, low-noise cryogenic resonance.*

The experimental setup involved positioning a dielectric target, a deionized water-filled sphere with a 40 mm diameter, at calibrated distances from the cryostat's optical window. The antenna's response was characterized by its reflection coefficient ($S_{11}$) at an input power of only 1 mW (0 dBm), a power level chosen to minimize any potential radiative heating of the cryogenic stage. The results, shown in Fig. 4, reveal a dual-modality sensing mechanism that operates in two distinct physical regimes: a near-field reactive regime dominated by frequency shifts, and a far-field radiative regime dominated by magnitude changes. We classify near/far field using the standard antenna criterion $r_{FF} \gtrsim 2D^2/\lambda$ (with $D$ the disk diameter); distances $d < r_{FF}$ are labeled near field in the text, while the panel labels ($\lambda/2, \lambda, 2\lambda, ...$) provide convenient normalized distances.

*1) Near-Field Reactive Sensing ($d \leq \lambda$)*

At distances close to an antenna, within approximately one wavelength ($\lambda$), the electromagnetic environment is dominated by the near field. Unlike the propagating far-field wave, the near field is a region of stored, non-propagating (evanescent) energy bound to the resonator. The strength of this reactive field decays rapidly with distance *r*, typically as $1/r^3$ or faster (*29*). When a dielectric object enters this region, it directly interacts with this stored energy, primarily perturbing the resonator's properties. This interaction is analogous to modifying a simple LC resonant circuit. The region of strong electric field in the resonator acts as its primary capacitance. Introducing a high-permittivity material like the water sphere ($\epsilon_r \approx 80$) into this E-field is equivalent to inserting a dielectric slab into a capacitor; it increases the system's ability to store electric energy, thereby increasing its effective capacitance, C. The resonant frequency, $f_0$, is inversely proportional to this capacitance ($f_0 \propto 1/\sqrt{LC}$). Consequently, an increase in capacitance must cause a decrease in the resonant frequency, resulting in a negative frequency shift.

This physical intuition is formalized by cavity perturbation theory. The corrected formula, derived from Slater's theorem, relates the frequency shift to the change in stored energy and explicitly accounts for this inverse relationship (*33*, *34*):

$$\frac{\Delta f_0}{f_0} \approx -\frac{\int_V (\Delta\epsilon|\vec{E_0}|^2 + \Delta\mu|\vec{H_0}|^2)dV}{2\int_V (\epsilon_0|\vec{E_0}|^2 + \mu_0|\vec{H_0}|^2)dV} \quad (3)$$

where $\vec{E_0}$ and $\vec{H_0}$ are the unperturbed fields, and $\Delta\epsilon$ is the change in permittivity. For a non-magnetic dielectric target ($\Delta\mu = 0$) with a permittivity greater than vacuum ($\Delta\epsilon > 0$), the integral in the numerator is positive. Due to the leading negative sign in (3), the resulting frequency shift, $\Delta f_0$, is negative. Our experimental measurements quantitatively confirm this model. We observed a maximum frequency shift of -2.5 MHz when the target was within one wavelength of the antenna. As expected, this frequency-shift-based sensing modality was highly sensitive only at this short range and became negligible for $d > 2\lambda$, consistent with the rapid decay of the near-field strength.

*2) Far-Field Radiative Sensing ($d > \lambda$)*

At larger distances ($d > \lambda$), the sensing mechanism transitions to a radiative regime. Here, the antenna's far-field wave illuminates the target, which then scatters a weak echo back toward the antenna. The total reflection coefficient measured by the network analyzer, $\Gamma_{\text{total}}$, is the coherent sum of the antenna's intrinsic reflection coefficient, $\Gamma_0$, and the complex reflection from the target, $\Gamma_{\text{target}}(d)$: $\Gamma_{\text{total}}(d) = \Gamma_0 + \Gamma_{\text{target}}(d)$. The target's reflection, $\Gamma_{\text{target}}(d)$, has a magnitude that decays with distance as $1/d^2$ (due to the round-trip path loss described by the radar range equation) and a phase that rotates with distance as $e^{-j2kd}$, where $k = 2\pi/\lambda$. This interference between a large, static vector ($\Gamma_0$) and a small, rotating vector ($\Gamma_{\text{target}}$) causes the magnitude of the $S_{11}$ dip to oscillate as the target distance changes.

In our data, moving the target from the AWO baseline ($\approx -35.95$ dB) to $d = \lambda/2$ deepens the dip to $\approx -43.02$ dB (a $\sim 7$ dB change), while other distances yield shallower or intermediate dips ($-34$ to $-36$ dB), consistent with constructive/destructive interference of the backscattered field with $\Gamma_0$. A detectable modulation persists out to $4\lambda$ ($\sim 46$ cm), despite the expected $1/d^4$ power scaling in a monostatic link. To isolate target-induced effects from window reflections and internal multipath, all far-field conclusions are referenced to the AWO baseline; the "window closed" trace (dip $\approx -22.43$ dB) illustrates the additional mismatch introduced by the sealed aperture and is not used for sensing metrics.

Together, the near-field frequency shifts and far-field amplitude modulations constitute a dual-modality, through-window sensing approach that operates at milliwatt-level input power and does not require any additional cabling into the 10 K stage. This establishes the feasibility of thermally efficient, cable-free diagnostics for cryogenic environments and provides a practical template for wireless interconnects in complex cryogenic systems.

## 3. CONCLUSIONS

This work has presented a direct, comparative analysis of MCT and ZST ceramic resonators, establishing a clear materials-based design framework for high-performance cryogenic microwave components. Under identical fixtures and a common calibration plane, and with both disks operated as dielectric resonator antennas, our results identify ZST as the superior material for applications demanding high stability and coherence. The ZST resonator demonstrated excellent thermal stability (average $\tau_f \approx -40.4$ ppm/K), a 22.6% enhancement in loaded Q-factor upon cooling to 10 K, and, critically, negligible thermal hysteresis. In stark contrast, the MCT composite, despite its stability at room temperature, proved unsuitable in its present composition/fixture for deep-cryogenic operation. It exhibited severe thermal instability ($\tau_f < -200$ ppm/K by magnitude), a collapse of its loaded Q to $\sim 7$ at 7 K, and substantial hysteresis (e.g., $\Delta f_0^{\text{hyst}} \approx 150$ MHz at 150 K), rendering it unreliable for any precision application. From on-resonance $|S_{11}|$ we further estimated the unloaded quality factor of ZST to increase from $\sim 48$ at 296 K to $\sim 64$ at 10 K, corroborating an intrinsic loss reduction upon cooling.

The implications are twofold. First, the demonstrated stability and Q-factor enhancement of ZST are advantageous for quantum readout architectures: for fixed external coupling, the higher $Q$ entails a proportionally smaller resonator decay rate $\kappa = 2\pi f_0/Q$, with our data indicating a linewidth reduction from $\kappa_L/2\pi \approx 93.5$ MHz (296 K) to $\approx 77.0$ MHz (10 K), i.e., $\sim 18\%$. Its frequency stability (low $|\tau_f|$) is likewise conducive to dense, multiplexed operation with reduced frequency collisions and crosstalk.

Second, we leveraged these properties to realize a through-window wireless link at 10 K using only 1 mW input power. Relative to the "antenna window open" baseline, the ZST antenna exhibited measurable near-field frequency shifts and far-field amplitude modulations in $S_{11}$, enabling target detection out to $4\lambda$. This proof-of-concept supports thermally efficient, cable-free interconnects that help alleviate the interconnect bottleneck in complex cryogenic systems.

In summary, this study provides not only quantitative performance benchmarks but also a clear physical explanation for the divergent behaviors of these twocritical ceramic systems at low temperatures. We establish $(Zr,Sn)TiO_4$ as a foundational material for the next generation of high-coherence quantum interfaces and wireless cryogenic sensors. Limitations of the present work include temperature-related material models in EM simulations and omission of the window in pattern calculations; nevertheless, experimental baselines (window closed/open) were used to isolate sensing effects. Future work will optimize ZST geometries and coupling for specific cQED use, explore processing routes to further elevate $Q_u$ by reducing extrinsic defects, and extending materials and antenna characterization into the sub-Kelvin regime, where continued materials engineering will be essential for pushing the frontiers of quantum information science (*35*).

## Acknowledgement

This research was supported by an appointment to the Intelligence Community Postdoctoral Research Fellowship Program at Florida International University, administered by Oak Ridge Institute for Science and Education (ORISE) through an interagency agreement between the U.S. Department of Energy and the Office of the Director of National Intelligence (ODNI). The authors thank the Department of Electrical and Computer Engineering (ECE) at Florida International University for providing access to facilities and resources that contributed to this work. Samples were prepared by T-Ceram, Czech Republic.

## Author Declarations

### Conflict of Interest

The authors have no conflicts to disclose.

### Author Contributions

**I. Torres:** Conceptualization (Supporting), Methodology (Equal), Software (Lead), Validation (Equal), Formal Analysis (Lead), Investigation (Equal), Data Curation (Lead), Writing – Original

Draft (Lead), Visualization (Lead). **A. Krasnok:** Conceptualization (Lead), Methodology (Equal), Formal Analysis (Supporting), Investigation (Equal), Resources (Lead), Writing – Review & Editing (Lead), Supervision (Lead), Project Administration (Lead), Funding Acquisition (Lead).

**Data Availability**

The data that support the findings of this study are available from the corresponding author upon reasonable request.